\documentclass[aps,prl,twocolumn,preprintnumbers,amsmath,amssymb,floatfix,longbibliography,superscriptaddress,10pt]{revtex4-2}

\usepackage{xcolor}
\usepackage{dsfont}
\usepackage{comment}
\usepackage{import}
\usepackage[colorlinks=true,allcolors=blue,urlcolor=blue]{hyperref}
\usepackage[sort&compress]{natbib}

\usepackage{amsthm}

\usepackage{amssymb} 
\usepackage[version=4]{mhchem} 

\usepackage{empheq} 
\usepackage{enumerate} 
\usepackage{subfiles} 

\usepackage[super]{nth}
\usepackage{comment} 
\usepackage{stackrel}
\usepackage{cancel} 

\usepackage{empheq} 

\newcommand{\acomm}[2]{\left\{#1,#2 \right\}}
\newcommand{\up}{\uparrow}
\newcommand{\down}{\downarrow}
\newcommand{\expect}[1]{\left\langle#1\right\rangle}

\newcommand{\vecg}[1]{\boldsymbol{#1}}
\newcommand{\vk}{\mathbf{k}}
\newcommand{\vkp}{{\mathbf{k}^\prime}}

\newcommand{\vq}{\mathbf{q}}

\newcommand{\vlambda}{\vecg{\lambda}}

\newcommand{\vDelta}{\vecg{\Delta}}

\renewcommand{\d}{\text{d}}

\newcommand{\partiald}[2]{\frac{\partial #1}{\partial #2}}

\newcommand{\totald}[2]{\frac{\d #1}{\d #2}}

\newcommand{\expb}[1]{\exp \left[#1\right]}

\DeclareMathOperator{\sech}{sech}

\DeclareMathOperator{\Tr}{Tr}

\DeclareMathOperator{\Ima}{Im}

\begin{document}
\title{Topological Landau Theory}

\author{Canon Sun}
\email{canon@ualberta.ca}
\affiliation{Department of Physics, University of Alberta, Edmonton, Alberta T6G 2E1, Canada}
\affiliation{Theoretical Physics Institute, University of Alberta, Edmonton, Alberta T6G 2E1, Canada}
\author{Joseph Maciejko}
\email{maciejko@ualberta.ca}
\affiliation{Department of Physics, University of Alberta, Edmonton, Alberta T6G 2E1, Canada}
\affiliation{Theoretical Physics Institute, University of Alberta, Edmonton, Alberta T6G 2E1, Canada}
\affiliation{Quantum Horizons Alberta, University of Alberta, Edmonton, Alberta T6G 2E1, Canada}

\begin{abstract}
We present an extension of Landau's theory of phase transitions by incorporating the topology of the order parameter. When the order parameter comprises several components arising from multiplicity in the same irreducible representation of symmetry, it can possess a nontrivial topology and acquire a Berry phase under the variation of thermodynamic parameters. To illustrate this idea, we investigate the superconducting phase transition of an electronic system with tetragonal symmetry and an attractive interaction involving two partial waves, both transforming in the trivial representation. By analyzing the time-dependent Ginzburg-Landau equation in the adiabatic limit, we show that the order parameter acquires a Berry phase after a cyclic evolution of parameters. We study two concrete models---one preserving time-reversal symmetry and one breaking it---and demonstrate that the nontrivial topology of the order parameter originates from thermodynamic analogs of gapless Dirac and Weyl points in the phase diagram. Finally, we identify an experimental signature of the topological Berry phase in a Josephson junction.
\end{abstract}
\maketitle

\makeatother
\emph{Introduction.}---Landau theory is a foundational framework for understanding phase transitions~\cite{Landau1965On}. Its central tenet is that a continuous phase transition can be succinctly described by an expansion of the free energy in powers of the order parameter, a quantity that distinguishes the symmetry-broken and unbroken phases. Owing to its generality, Landau theory provides important insights into the critical phenomena of a vast array of physical systems, such as ferromagnets, antiferromagnets, Bose-Einstein condensates, superfluids, superconductors, and liquid crystals~\cite{Landau2013Statistical,Landau1980Statistical}. Furthermore, it can be applied to understand the physics of the ordered phases themselves. For instance, Ginzburg-Landau theory, an extension of Landau theory that incorporates spatial variations in the order parameter, can explain key phenomena in superconductivity such as dissipationless currents, the Meissner effect, and magnetic vortices~\cite{Ginzburg1965On}. As a result, Landau theory remains an indispensable tool for understanding various complex orders and their critical behavior.

Symmetry plays a crucial role in Landau theory~\cite{Dresselhaus2007Group}. As the free energy must respect the symmetries of the system, its expansion in powers of the order parameter must do so as well, which allows it to be constructed systematically. At each order in the series, one includes all invariants that remain unchanged under every symmetry operation. Among these terms, the second-order contribution is arguably the most important, as it determines the critical temperature and the relevant order parameter. The structure of the quadratic term ensures that order parameters transforming under different irreducible representations (irreps) of the symmetry group generally have distinct critical temperatures. Consequently, close to the phase transition, only the irrep with the highest transition temperature is relevant. In the case of superconductors, this observation underpins the established doctrine that superconducting states are to be classified according to their pairing symmetry~\cite{Sigrist1991Phenomenological}.

\begin{figure}[t!]
\includegraphics[width=0.98\columnwidth]{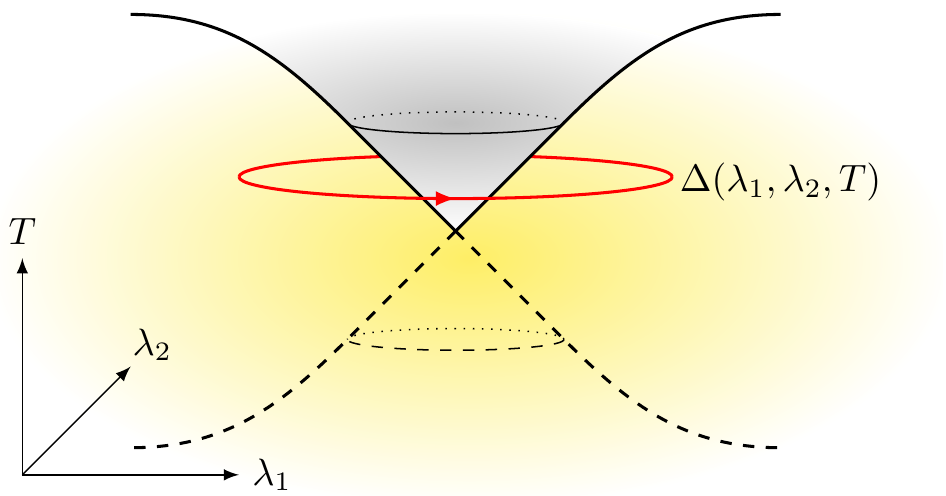}
\caption{\label{fig: schematic diagram} Schematic representation of the thermodynamic phase diagram considered in this work. Two critical surfaces (solid vs dotted lines) touch at a gapless Dirac point in the (2+1)-dimensional parameter space defined by interaction parameters $\lambda_1$, $\lambda_2$ and temperature $T$. The symmetry-broken and unbroken phases, distinguished by the order parameter $\Delta(\lambda_1,\lambda_2,T)$, occupy the yellow and gray regions, respectively. When $\lambda_1$ and $\lambda_2$ are varied adiabatically in a closed loop, the order parameter $\Delta$ may acquire a nontrivial Berry phase.}
\end{figure}

In this Letter, we study the Landau theory of two ``competing'' orders that transform under the same irrep~\footnote{While in Landau theory, ``competing orders'' typically refers to symmetry-distinct orders $\phi_1,\phi_2$ with a (repulsive) biquadratic coupling $\sim|\phi_1|^2|\phi_2|^2$ in the free energy~\cite{KivelsonStatMech}, here we mean distinct but symmetry-compatible orders that are allowed to mix at the bilinear (quadratic) level.} and investigate the topology of the resulting order parameter.  The central observation is that when the order parameter comprises multiple components that transform under the same irrep, the quadratic term in the Landau free energy assumes a matrix structure. The thermodynamically stabilized order parameter, which is an eigenvector of this quadratic matrix $A$, represents a mixture of the two competing orders. Generally, $A$ depends on a collection of parameters $\vlambda$, which may be of microscopic origin, such as the interaction strength, or macroscopic, e.g. temperature. When these parameters are varied adiabatically, the order parameter remains in its instantaneous free-energy minimum. As a consequence, if the parameters are varied in a closed loop, the order parameter can acquire a Berry phase, as illustrated in Fig.~\ref{fig: schematic diagram}, revealing its nontrivial topology. The analysis of the topology of the order parameter bears certain parallels with topological band theory~\cite{Hasan2010Colloquium,Qi2011Topological,Chiu2016Classification}. In this analogy, the parameters $\vlambda$ play the role of the crystal momentum, the matrix $A$ replaces the Bloch Hamiltonian, the order parameter substitutes the Bloch state, and the critical surfaces in the phase diagram correspond to electronic bands, as shown in Fig.~\ref{fig: schematic diagram}. Because of the similarities, we refer to this generalization of Landau theory as \emph{topological Landau theory}.

To illustrate the above ideas, we focus on a superconductor with tetragonal symmetry (point group $D_{4h}$) and an attractive interaction involving two partial waves that transform in the same irrep. We begin by analyzing the Landau free energy to identify the static thermodynamic ground state. Next, we explore the adiabatic dynamics, demonstrating that the Berry phase arises naturally from the time-dependent Ginzburg-Landau (TDGL) equation~\cite{Landau1954On, Schmid1966Time, Abrahams1966Time, Gorkov1968Generalization, Stoof1993Time} in the adiabatic limit. We then examine the topology of the order parameter in two specific models: one preserving time-reversal symmetry and the other breaking it. Finally, we discuss how the Berry phase can be experimentally detected in a Josephson junction and comment on extensions to other thermodynamic systems and cases involving higher-dimensional irreps.

\emph{Landau free energy.}---Consider a three-dimensional (3D) superconductor with $D_{4h}$ symmetry near the metal--superconductor transition. The system is described by the Hamiltonian 
\begin{equation}\label{eq: Ham}
    \hat{H}=\hat{H}_0+\hat{H}_{\text{int}},
\end{equation}
where 
\begin{align}
    \hat{H}_0&= \sum_{\vk\sigma}\xi_\vk \hat{c}^\dagger_{\vk \sigma} \hat{c}_{\vk\sigma},\\
    \hat{H}_{\text{int}}&=-\frac{1}{\mathcal{V}}\sum_{\vk\vkp}V_{\vk\vkp}\hat{c}^\dagger_{\vk\up}\hat{c}^\dagger_{-\vk\down}\hat{c}_{-\vkp\down}\hat{c}_{\vkp\up}.
\end{align}
Here $\hat{c}^\dagger_{\vk\sigma}$ is the creation operator for an electron with wavevector $\vk$ and spin $\sigma=\up,\down$ and $\mathcal{V}$ is the volume of the system. The first term is the free-electron Hamiltonian 
with single-particle dispersion $\xi_\vk=\hbar^2k^2/(2m)-\mu$ and the second is an interaction Hamiltonian that describes the scattering of Cooper pairs in the singlet channel. The interaction matrix element is given by
\begin{equation}\label{eq: partial wave decomposition}
V_{\vk\vkp}=\sum_{\alpha,\beta=1}^2V_{\alpha\beta}\phi_\alpha(\vk)\phi_\beta(\vkp),
\end{equation}
where $\phi_1(\vk)=1$ and $\phi_2(\vk)= \sqrt{5}[1-3(k_x^2+k_y^2)/(2k_F^2)]$ are $D_{4h}$ partial waves transforming in the trivial representation, normalized to $\int \d\Omega_\vk \phi_\alpha(\vk)\phi_\beta(\vk)/(4\pi)=\delta_{\alpha\beta}$. The interaction matrix elements $V_{\alpha\beta}$ form a complex Hermitian matrix $V$, whose precise form will be specified later. We assume $V$ is positive-definite so that the interaction is attractive. The time-reversal operator acting on $V$ is $\hat{\mathcal{T}}= \hat{K}$, where $\hat{K}$ is the complex-conjugation operator, and satisfies $\hat{\mathcal{T}}^2=1$. Time-reversal symmetry is preserved if and only if $V$ is real.

The superconducting phase transition can be captured by the Landau free energy, with the gap function serving as the order parameter. Since the interaction potential Eq.~\eqref{eq: partial wave decomposition} consists of only the partial waves $\phi_1(\vk)$ and $\phi_2(\vk)$, near the phase transition, the gap function adopts the general form
\begin{equation}\label{eq: Delta partial wave decomposition}
    \Delta_\vk = \sum_{\alpha=1}^2 \Delta_\alpha \phi_\alpha(\vk),
\end{equation}
where $\Delta_\alpha$ are complex amplitudes. In other words, the superconducting phase is characterized by a two-component order parameter, $\vDelta=(\Delta_1,\Delta_2)^T$. To fourth order, the mean-field free energy density has the form 
\begin{equation}\label{eq: Landau free energy}
    f(\vecg{\Delta})\simeq \sum_{\alpha\beta}\Delta_\alpha^*A_{\alpha\beta}\Delta_\beta+ \frac{1}{2}\sum_{\alpha\beta\gamma\delta}B_{\alpha\beta\gamma\delta}\Delta_\alpha^*\Delta_\beta^*\Delta_\gamma\Delta_\delta.
\end{equation}
The quadratic coefficients are given by
\begin{equation}
A_{\alpha\beta}=\left[V^{-1}-N(0)\ln \left(\frac{2e^\gamma}{\pi}\frac{\Lambda}{k_BT} \right)I_2\right]_{\alpha\beta},
\end{equation}
where $N(0)= (2m/\hbar^2)^{3/2}\sqrt{\mu}/(4\pi^2)$ is the density of states per spin at the Fermi level, $\gamma$ is the Euler-Mascheroni constant, $\Lambda$ is the energy cutoff for the interaction, and $I_2$ is the $2\times 2$ identity matrix. Crucially, the order parameters $\Delta_1$ and $\Delta_2$ are allowed to couple at the quadratic level because $\phi_1(\vk)$ and $\phi_2(\vk)$ transform in the same irrep, which would otherwise be prohibited by symmetry. The quartic coefficients are given by
\begin{equation}
    B_{\alpha\beta\gamma\delta}=\frac{7\zeta(3)N(0)}{8\pi^2( k_BT)^2}\int \frac{\d\Omega_\vk}{4\pi}\phi^*_\alpha(\vk)\phi_\beta^*(\vk) \phi_\gamma(\vk)\phi_\delta(\vk),
\end{equation}
where $\zeta(z)$ is the Riemann zeta function. A detailed derivation of the Landau free energy can be found in Supplementary Material (SM) I. 

Near the transition, the form of the order parameter is completely determined by the quadratic term in the free energy.  Let $\hat{\vecg{\Delta}}_{\pm}$ be normalized eigenvectors of $A$ with eigenvalues $a_{\pm}$, where $a_+>a_-$, and some fixed choice of phase. When $a_->0$, the minimum of the free energy is at $\vecg{\Delta}=0$ and the system is in the metallic state. As $a_-$ passes from positive to negative, the $\hat{\vDelta}_-$ component condenses and the system transitions into the superconducting state. Thus, close to the critical point, the order parameter has the form $\vecg{\Delta}=\Delta_0 e^{i\varphi}\hat{\vecg{\Delta}}_-$.
The gap magnitude is determined by minimizing the full Landau free energy and is given by $\Delta_0=\sqrt{-a_-/b_-}$, where $b_-=\sum_{\alpha\beta\gamma\delta}B_{\alpha\beta\gamma\delta}\hat{\Delta}_{-,\alpha}^*\hat{\Delta}_{-,\beta}^*\hat{\Delta}_{-,\gamma}\hat{\Delta}_{-,\delta}$. The phase $e^{i\varphi}$ is spontaneously chosen and is defined relative to the phase convention of $\hat{\vecg{\Delta}}_-$. Importantly, the contribution from the high-energy component $\hat{\vecg{\Delta}}_+$ can be ignored when $a_+\gg |B_{\alpha\beta\gamma\delta}||\Delta_0|^2$ for all indices $\alpha,\beta,\gamma,\delta$, which is the case sufficiently close to the phase transition.

\emph{Adiabatic dynamics.}---We now examine the adiabatic dynamics of the order parameter by varying the interaction matrix elements $ V_{\alpha\beta}$ continuously. Suppose the interaction matrix elements $V_{\alpha\beta}=V_{\alpha\beta}(\vlambda)$ depend on a collection of parameters $\lambda_j$, $j=1,\dots,N$, which are slowly varied in time. The low-frequency dynamics of the superconducting gap function near the phase transition is governed by the TDGL equation~\cite{Landau1954On, Schmid1966Time, Abrahams1966Time, Gorkov1968Generalization, Stoof1993Time}. For the model described in Eq.~\eqref{eq: Ham}, the TDGL equation takes the form 
\begin{equation}\label{eq: TDGL}
 -  \hbar\eta\totald{\Delta_\alpha}{t}= \partiald{f}{\Delta_\alpha^*},
 \end{equation}
 where $\eta=\pi N(0)/(8k_BT)$. A detailed derivation of Eq.~\eqref{eq: TDGL} can be found in SM II.

There are three timescales pertinent to the dynamics. The times $\tau_{\pm}\equiv \hbar\eta/|a_\pm|$ are the characteristic timescales at which deviations of the gap function from its equilibrium value in the $\hat{\vDelta}_{\pm}$ directions relax. By assumption, $\tau_+\ll \tau_-$ near the phase transition. We also define $\tau$ to be the characteristic timescale at which the parameters $\vlambda$ are varied. Although $\tau_-$ diverges at the critical point, it is typically small slightly away from the phase transition. Thus, we have the hierarchy of scales $\tau_+\ll \tau_-\ll \tau$ and can treat the system as quasistatic, in which case the gap function takes the form
\begin{equation}\label{eq: gap equilibrium}
    \vDelta(\vlambda)=\Delta_0(\vlambda)e^{i\varphi(t)}\hat{\vDelta}_-(\vlambda),
\end{equation}
where $\hat{\vDelta}_-(\vlambda)$ is an instantaneous eigenvector of $A(\vlambda)$ with some fixed choice of phase. The gap magnitude $\Delta_0(\vlambda)$ minimizes the free energy at $\vlambda$ and the phase $e^{i\varphi(t)}$ is defined with respect to $\hat{\vDelta}_-(\vlambda)$.

In the quasistatic regime, while the magnitude $\Delta_0(\vlambda)$ and basis state $\hat{\vDelta}_-(\vlambda)$ are determined once the parameters $\vlambda$ are specified, the phase $\varphi$ is not and can depend on the path taken in parameter space. As shown in SM III, the phase $\varphi$ remains unchanged under relaxation, indicating that any variation arises solely from the adiabatic evolution of $\vlambda$. Assuming the system does not undergo a phase transition as the parameters are varied, in which case the order parameter $\vecg{\Delta}(\vlambda)$ depends smoothly on the parameters $\vlambda$, we substitute the ansatz~\eqref{eq: gap equilibrium} into the TDGL equation to obtain
\begin{align}
    \totald{\Delta_0}{t}(\vlambda(t))&=0,\label{eq: magnitude EOM}\\
  \totald{\varphi}{t}(t)&= \sum_j\mathcal{A}_{-,j}(\vlambda) \totald{\lambda_j}{t},\label{eq: phase EOM}
\end{align}
where $\mathcal{A}_{-,j}(\vlambda)=i\hat{\vecg{\Delta}}^\dagger_-(\vlambda)\partial_j \hat{\vDelta}_-(\vlambda)$ is the abelian Berry connection associated with the pairing state $\hat{\vDelta}_-(\vlambda)$. In reaching Eq.~\eqref{eq: phase EOM}, we have ignored the contribution from the nonabelian Berry connection  $i\hat{\vDelta}_+^\dagger(\vlambda)\partial_j\hat{\vDelta}_-(\vlambda)$, which is negligible when $\tau$ is much greater than $\hbar\eta/(a_+-a_-)\simeq \tau_+$. The equation of motion for the magnitude, Eq.~\eqref{eq: magnitude EOM}, is automatically satisfied in the adiabatic limit. For the phase, if the parameters are varied in a closed loop $C$ in parameter space, it becomes
\begin{equation}
    \varphi=\varphi_0+\oint_C \sum_j\d\lambda_j \mathcal{A}_{-,j}(\vlambda),
\end{equation}
where $\varphi_0$ is the initial phase. Thus, under a cyclic, adiabatic evolution of parameters, the gap function acquires a geometric phase determined by the connection $\mathcal{A}_{-,j}$. 

\begin{figure}[t!]
\includegraphics[width=0.98\columnwidth]{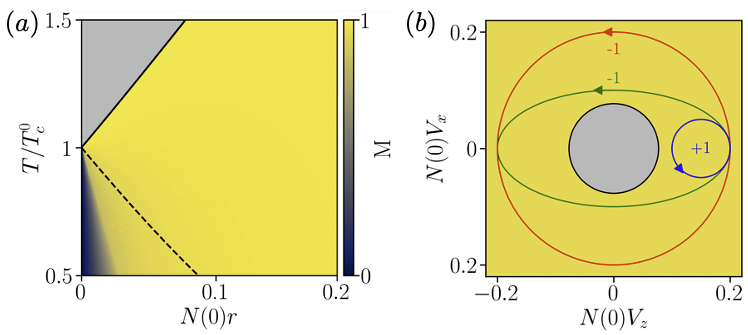}
\caption{\label{fig: phase diagram} (a) Phase diagram of the model in Eq.~\eqref{eq: TRI model}, which features the thermodynamic analog of a (2+1)D Dirac point at $r=0$, $T=T_c^0$ (gray: metallic state; remaining regions: superconducting state; solid and dashed lines: critical lines defined by $a_-=0$ and $a_+=0$, respectively). The colorbar measures $M= |\hat{\vDelta}^\dagger_-\cdot \hat{\vDelta}|^2$, the squared overlap of the normalized numerical ground state $\hat{\vDelta}$ with the state $\hat{\vDelta}_-$. Here $k_B T_c^0 = 2\Lambda e^\gamma e^{-1/(N(0)V_0)}/\pi$ is the critical temperature when $r=0$. Parameter values are $\Lambda=10$, $N(0)V_0=0.4$, and $\phi=2.7$. (b) Schematic view of a constant-temperature $(T>T_c^0)$ slice of the phase diagram. The axes correspond to $N(0)V_z=N(0)r\cos\phi$ and $N(0)V_x= N(0)r\sin\phi$. The order parameter acquires a minus sign after traversing a closed path encircling the metallic part of the phase diagram once (red and green loops). Conversely, paths that do not encircle the origin (blue) result in no Berry phase.}
\end{figure}

\emph{Dirac and Weyl points in Landau theory.}---To illustrate the general ideas discussed above, we examine two concrete models with distinct interaction matrices, $V$. First, we consider the most general interaction matrix compatible with time-reversal symmetry, which gives the thermodynamic analog of a (2+1)D Dirac point:
\begin{equation}\label{eq: TRI model}
    V= V_0I_2 + r(\cos\phi\sigma_z+\sin\phi\sigma_x),
\end{equation}
where $e^{i\phi}\in S^1$ and $V_0>r>0$ to ensure both partial wave channels are attractive and $\sigma_{x,y,z}$ are the Pauli matrices. Because of time-reversal symmetry, the eigenvectors of the associated quadratic matrix $A(r,\phi)$ can be chosen to be real at all parameter values. The eigenvectors with lower and higher critical temperatures are, respectively, $\hat{\vDelta}_-(r,\phi)=(\cos(\phi/2),\sin(\phi/2))^T$ and $\hat{\vDelta}_+(r,\phi)=(-\sin(\phi/2),\cos(\phi/2))^T$.

To verify the predictions of Landau theory, we determine the phase diagram by minimizing the full mean-field free energy (see SM I for the explicit expression). The resulting phase diagram is shown in Fig.~\ref{fig: phase diagram}(a). At each parameter value, we compute $M =|\hat{\vDelta}_-^\dagger(r,\phi)\cdot\hat{\vDelta}(r,\phi)|^2$, which measures the squared overlap between the normalized numerically computed ground state, $\hat{\vDelta}(r,\phi)$, and the prediction from the quadratic term, $\hat{\vDelta}_-(r,\phi)$. Near the phase transition from the normal state, we find $M\approx 1$, in agreement with the predictions of Landau theory.

The pairing state $\hat{\vDelta}_-(r,\phi)$ can exhibit a nontrivial Berry phase as the parameters $r$ and $\phi$ are varied. Although the Berry connection vanishes for all $\phi\in[0,2\pi)$, there can still be a Berry phase due to a mismatch between the initial and final states~\cite{Vanderbilt2018Berry}. Indeed, the order parameter is not single valued: $\hat{\vDelta}_-(r,\phi)= -\hat{\vDelta}_-(r,\phi+2\pi)$. Therefore, for a loop in parameter space that encircles the Dirac point at $r=0$, the gap function acquires a $\pi$ Berry phase. On the other hand, if the loop does not enclose the origin, it does not pick up a Berry phase. This is illustrated schematically in Fig.~\ref{fig: phase diagram}(b).

As the second example, we consider the most general interaction matrix element possible, which breaks time-reversal symmetry and gives the thermodynamic analog of a (3+1)D Weyl point:
\begin{equation}
    V=V_0I_2+ r\left(\cos\theta\sigma_z+\sin\theta(\cos\phi \sigma_x+\sin\phi\sigma_y) \right).
\end{equation} 
Here $(\theta,\phi)$ are the usual polar and azimuthal angles on the sphere and $V_0>r>0$. The eigenvectors of the matrix $A(r,\phi)$ are $\hat{\vecg{\Delta}}_-(\theta,\phi) = (\cos(\theta/2),\sin(\theta/2)e^{i\phi})^T$ and $\hat{\vecg{\Delta}}_+(\theta,\phi) = (-\sin(\theta/2)e^{-i\phi},\cos(\theta/2))^T$. The more attractive partial-wave channel $\hat{\vDelta}_-(\theta,\phi)$ has a lower transition temperature and is thus stabilized over $\hat{\vDelta}_+(\theta,\phi)$.

The topology of the gap function is manifested through its Berry phase structure. The Berry connection of the state $\hat{\vDelta}_-(\theta,\phi)$, treating $\theta$ and $\phi$ as the parameters, is $\mathcal{A}_{-,\theta}=0$ and $\mathcal{A}_{-,\phi}=2q \sin^2(\theta/2)$, where $q=-1/2$ is the monopole charge. The associated Berry curvature is $\mathcal{F}_{-,\theta\phi}\equiv\partial_{\theta}\mathcal{A}_{-,\phi}-\partial_\phi \mathcal{A}_{-,\theta}=q\sin\theta$, which is the curvature of a magnetic monopole with monopole charge $q$ at $r=0$. For a loop in parameter space that subtends solid angle $\Omega$ about the Weyl point at $r=0$, the Berry phase acquired by the order parameter is $e^{iq\Omega}$. We note that the monopole here is in the parameter space defined by the interaction matrix elements. This is in contrast to monopole superconductivity, which also features a topological order parameter and a nontrivial pair Berry phase, but with the monopole situated in momentum space~\cite{Murakami2003Berry, Li2018Topological, Sun2019Vortices, Sun2020Z2, Bobrow2022Monopole, Li2024Berry, Frazier2024Designing, Bobrow2020Monopole}.

\begin{figure}[t]
\includegraphics[width=0.9\columnwidth]{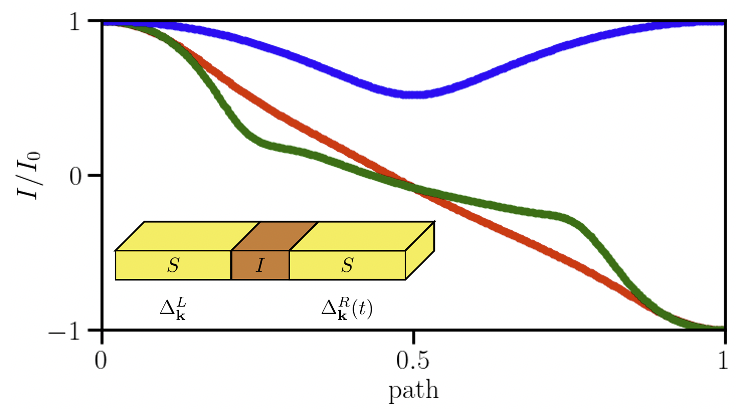}
\caption{\label{fig: Josephson} 
Normalized Josephson current, $I/I_0$, across the junction as the parameters in the right superconductor are evolved along three different paths in parameter space. Here, $I_0$ represents the current for the initial configuration 
$\Delta_\vk^L(N(0)r=0.2,\phi=0)=-i\Delta_\vk^R(N(0)r=0.2,\phi=0)$. The three curves correspond to the paths of matching color shown in Fig.~\ref{fig: phase diagram}(b). Inset: schematic of a Josephson junction.}
\end{figure}

\emph{Topological Josephson effect.}---The Berry phase of the gap function acquired from adiabatic evolution can be measured by the Josephson effect. Consider a system consisting of two superconductors separated by a weak link, as shown in the inset of Fig.~\ref{fig: Josephson}. The gap functions on the left and right superconductors are denoted as $\Delta^L_\vk$ and $\Delta^R_\vk$, respectively. To describe the coupling between the superconductors, we introduce a tunneling Hamiltonian of the form $\hat{H}_T=t\sum_{\vk\vq,\sigma}\hat{c}^{L\dagger}_{\vk\sigma}\hat{c}^R_{\vq\sigma}+\text{h.c.}$, where $t$ is the tunneling amplitude and $\hat{c}^L_{\vk\sigma}$ and $\hat{c}^R_{\vq\sigma}$ are electron annihilation operators for the superconductors on the left and right, respectively. This tunneling leads to a Josephson current across the junction, given by~\cite{Abrikosov2017Fundamentals}
\begin{widetext}
\begin{align}\label{eq: Josephson current}
    I_J=\frac{2e|t|^2}{\hbar}[N(0)]^2\mathcal{V}_L\mathcal{V}_R\Ima \int \frac{\d\Omega_\vk}{4\pi}\frac{\d\Omega_\vq}{4\pi}\int_{-\Lambda}^\Lambda \d \xi_\vk \d\xi_\vq \frac{\Delta^L_\vk \Delta_\vq^{R*}}{E_\vk^LE_\vq^R}\left[\frac{n_{\text{F}}(E_\vk^L)-n_{\text{F}}(E_\vq^R)}{E_\vk^L-E_\vq^R}+ \frac{1-n_{\text{F}}(E_\vk^L)-n_{\text{F}}(E^R_\vq)}{E^L_\vk+E^R_\vq} \right],
\end{align}
\end{widetext}
where $n_{\text{F}}(\xi)$ is the Fermi-Dirac distribution, $E_\vk^L = \sqrt{\xi_\vk^2 + |\Delta_\vk^L|^2}$ is the quasiparticle energy in the left superconductor and $\mathcal{V}_L$ is its volume (similarly for the right superconductor).

To reveal the nontrivial Berry phase, we compute the Josephson current across the junction as the parameters in the right superconductor are varied adiabatically, while those on the left are held fixed. For the time-reversal symmetric model, starting from a configuration satisfying $\Delta^L_\vk = -i\Delta_\vk^R$, we evaluate the Josephson current numerically using Eq.~\eqref{eq: Josephson current} at each point in the adiabatic evolution for the three paths in Fig.~\ref{fig: phase diagram}(b). As shown in Fig.~\ref{fig: Josephson}, the Josephson current for the topologically nontrivial paths (red and green) switch direction after a cycle due to the Berry phase. In contrast, for the contractible path (blue), the current returns to its original value.

\emph{Generalization to other broken symmetries.}---The above ideas can be generalized to other symmetry-breaking transitions. Consider a generic thermodynamic system in the vicinity of a continuous phase transition. Suppose the symmetry-broken and unbroken phases are distinguished by an order parameter $\Psi$ that resides in some Hilbert space $\mathcal{H}$. Let $G$ be the symmetry group of the system. The group $G$ is represented unitarily on $\mathcal{H}$, allowing $\mathcal{H}$ to be decomposed into invariant subspaces of $G$: $\mathcal{H}=\bigoplus_{\mu,\alpha}\mathcal{H}^{(\mu)}_\alpha$, where $\mu$ labels the irreps of $G$ and $\alpha$ the different orthogonal subspaces that all transform in the irrep $\mu$. We denote the components of $\Psi$ in $\mathcal{H}^{(\mu)}_\alpha$ relative to some orthonormal basis as $\Psi^{(\mu)}_{\alpha m}$, where $m=1,\dots, d_\mu$ labels the vectors transforming in the irrep $\mu$ and $d_\mu$ is the dimension of $\mu$. 

The relevant order parameter is determined by the quadratic term in the free energy. The most general quadratic contribution compatible with symmetry takes the form
\begin{equation}
f_2(\Psi)=\sum_{\mu,\alpha\beta,m}\Psi^{(\mu)*}_{\alpha m}A^{(\mu)}_{\alpha\beta}(\vlambda)\Psi^{(\mu)}_{\beta m},
\end{equation}
where $A^{(\mu)}(\vlambda)$ is a Hermitian matrix that depends smoothly on a collection of parameters $\vlambda$. In this more general scenario, the quadratic matrix $A=\oplus_\mu A^{(\mu)}$ is a direct sum of smaller matrices $A^{(\mu)}$, which are labelled by their irrep. As before, the order parameter is the eigenvector of $A$ with the smallest eigenvalue, which transforms in a particular irrep. Sufficiently near the transition, the order parameter takes the form $\Psi(\vlambda) = \sum_m c_m(\vlambda)\hat{\Psi}^{(\mu)}_{-, m}(\vlambda)$, where $\hat{\Psi}^{(\mu)}_{-,m}(\vlambda)$, with $m=1,\dots,d_\mu$, are normalized eigenvectors of $A$ associated with the smallest eigenvalue, and $c_m(\vlambda)$ are coefficients that are determined by minimizing the full Landau free energy. The dependence of the eigenvectors $\hat{\Psi}^{(\mu)}_{-, m}(\vlambda)$ on the parameters $\vlambda$ can give rise to a geometric contribution to the adiabatic equations of motion.

\emph{Summary and outlook.}---In conclusion, we have analyzed the Landau theory of two superconducting orders with the same pairing symmetry and explored the topology of the resulting order parameter. The stabilized order parameter is a mixture of the two orders and, as the interaction parameters are varied, the order parameter can acquire a Berry phase. The origin of this Berry phase lies in the presence of diabolical points, such as Dirac and Weyl points, in the thermodynamic phase diagram. Finally, we showed that the Berry phase can be experimentally measured in a Josephson junction.

In this work, we highlighted the foundational ideas of topological Landau theory, laying the groundwork for future studies. On the theoretical front, it would be interesting to investigate systems with a nonabelian Berry connection and extend beyond superconductivity. It would also be intriguing to measure the Berry phase experimentally. In particular, ultracold atoms, where the interaction strength can be tuned via Feshbach resonances~\cite{Barankov2004Collective, Barankov2006Synchronization, Barankov2006Dynamical,Bohn2000Cooper, Holland2001Resonance, Kokkelmans2002Resonance,Milstein2002Resonance,Ohashi2002BCS, Ohashi2003Superfluid, Ohashi2003Superfluidity,Stajic2004Nature}, provide a promising platform to observe this Berry phase. 

\emph{Acknowledgements.}---The authors thank Albion Arifi, Igor Boettcher, Davidson Noby Joseph, Matthias Kaminski, Hanbang Lu, Frank Marsiglio, Leyla Saraj, and Connor Walsh for fruitful discussions. C.S. acknowledges support through the Natural Sciences and Engineering Research Council of Canada (NSERC) Discovery Grant RGPAS-2020-00064 and the Pacific Institute for the Mathematical Sciences CRG PDF Fellowship Award. 
J.M. was supported by NSERC Discovery Grants RGPIN-2020-06999 and
RGPAS-2020-00064; the Canada Research Chair (CRC) Program; and Alberta Innovates.

\clearpage
\onecolumngrid
\appendix

\section{Supplemental Material for ``Topological Landau Theory''} 
\renewcommand{\theequation}{S\arabic{equation}}
\setcounter{equation}{0}  

\subsection{I. Mean-field free energy} 
\label{{app: MF free energy}}
In this supplementary material, we derive the mean-field free energy used to obtain the phase diagram in Fig.~\ref{fig: phase diagram}(a). From this, we construct the Landau free energy, Eq.~\eqref{eq: Landau free energy}, by performing an expansion to fourth order in the gap coefficients $\Delta_\alpha$.

In the superconducting state, the pair amplitude $\expect{\hat{c}_{-\vkp\down}\hat{c}_{\vkp\up}}$ becomes nonzero because of the condensation of Cooper pairs. In the mean-field approximation, the Hamiltonian~\eqref{eq: Ham} becomes
\begin{equation}
    \hat{H}=\hat{H}_0- \sum_\vk\left(\Delta_\vk\hat{c}^\dagger_{\vk\up}\hat{c}^\dagger_{-\vk\down}+\text{h.c.}\right)+K_0,
\end{equation}
where 
\begin{equation}\label{eq: app gap function definition}
    \Delta_\vk = \frac{1}{\mathcal{V}}\sum_{\vkp}V_{\vk\vkp}\expect{\hat{c}_{-\vkp\down}\hat{c}_{\vkp\up}}, 
\end{equation}
is the microscopic definition of the gap function, and
\begin{align}
    K_0= \frac{1}{\mathcal{V}}\sum_{\vk}V_{\vk\vkp}\expect{\hat{c}^\dagger_{\vk\up}\hat{c}^\dagger_{-\vk\down}}\expect{\hat{c}_{-\vkp\down}\hat{c}_{\vkp \up}   },
\end{align}
is a constant shift in the energy. Let us first write the constant shift $K_0$ in terms of the order parameter coefficients $\Delta_\alpha$. Using the partial wave decompositions Eqs.~\eqref{eq: partial wave decomposition} and \eqref{eq: Delta partial wave decomposition}, we have
\begin{equation}
    \Delta_\alpha =\frac{1}{\mathcal{V}}\sum_{\vkp}\sum_\beta V_{\alpha\beta}\phi_\beta^*(\vkp)\expect{\hat{c}_{-\vkp\down}\hat{c}_{\vkp\up}}.
\end{equation}
This allows us to write $K_0$ as
\begin{align}
    K_0&= \frac{1}{\mathcal{V}}\sum_{\vk\vkp}\sum_{\alpha\beta}V_{\alpha\beta}\phi_\alpha(\vk)\phi_\beta^*(\vk)\expect{\hat{c}^\dagger_{\vk\up}\hat{c}^\dagger_{-\vk\down}}\expect{\hat{c}_{-\vkp\down}\hat{c}_{\vkp\up}}\nonumber\\
    &=\mathcal{V}\sum_{\gamma\delta}\left[\frac{1}{\mathcal{V}}\sum_\vk \sum_{\alpha}V_{\gamma\alpha}^*\phi_\alpha(\vk)\expect{\hat{c}^\dagger_{\vk\up}\hat{c}^\dagger_{-\vk\down}} \right]V^{-1}_{\gamma\delta}\left[ \frac{1}{\mathcal{V}}\sum_{\vkp}\sum_{\beta}V_{\delta\beta}\phi_\beta^*(\vkp) \expect{\hat{c}_{-\vkp\down}\hat{c}_{\vkp\up}} \right]\nonumber\\
    &=\mathcal{V}\sum_{\gamma\delta}\Delta^*_\gamma V_{\gamma\delta}^{-1} \Delta_\delta.
    \end{align}
In the second step, we have used that $V_{\alpha\beta}$ is a Hermitian matrix and is invertible because it is positive definite. 

The excitations in a superconductor are described by non-interacting Bogoliubov quasiparticles. Using the fermion anticommutation relations, the Hamiltonian can be brought to Nambu form:
\begin{align}
\hat{H}&=\sum_{\vk}\begin{pmatrix}
    \hat{c}^\dagger_{\vk\up}&\hat{c}_{-\vk\down}
\end{pmatrix}
\begin{pmatrix}
    \xi_\vk&-\Delta_\vk\\
    -\Delta_\vk^*&-\xi_\vk
\end{pmatrix}
\begin{pmatrix}
    \hat{c}_{\vk\up}\\\hat{c}_{-\vk\down}^\dagger
\end{pmatrix}+\sum_\vk \xi_\vk+K_0.
\end{align}
As the Hamiltonian is quadratic in fermion operators, it can be diagonalized by a Bogoliubov transformation. The transformation is given by
\begin{equation}
\begin{pmatrix}
    \hat{c}_{\vk\up}\\
    \hat{c}^\dagger_{-\vk\down}
\end{pmatrix}
= \begin{pmatrix}
    u_\vk^*&v_\vk\\
    -v_\vk^*&u_\vk
\end{pmatrix}
\begin{pmatrix}
    \hat{\gamma}_{\vk\up}\\
    \hat{\gamma}^\dagger_{-\vk\down}
\end{pmatrix},
\end{equation}
where 
\begin{align}\label{eq: coherence factors}
    u_\vk &=\frac{E_\vk+\xi_\vk}{\sqrt{2E_\vk(E_\vk+\xi_\vk)}},& v_\vk&=\frac{\Delta_\vk}{\sqrt{2E_\vk(E_\vk+\xi_\vk)}},
\end{align}
are the coherence factors and $E_{\vk}=\sqrt{\xi_\vk^2+|\Delta_\vk|^2}$ is the quasiparticle energy. The transformation brings the Hamiltonian to the form
\begin{equation}\label{eq: app H gamma}
    \hat{H}= \sum_{\vk\sigma}E_\vk\hat{\gamma}^\dagger_{\vk\sigma}\hat{\gamma}_{\vk\sigma}+\sum_\vk(\xi_\vk-E_\vk)+K_0.
\end{equation}
Since the transformation is unitary, the Bogoliubov quasiparticle operators obey the canonical anticommutation relations $\acomm{\hat{\gamma}_{\vk\sigma}}{\hat{\gamma}^\dagger_{\vkp\sigma^\prime}}=\delta_{\vk\vkp}\delta_{\sigma\sigma^\prime}$. Thus, the Hamiltonian Eq.~\eqref{eq: app H gamma} describes a collection of free fermions, with additional constant terms.

The thermodynamic ground state is obtained by minimizing the free energy. Given the Hamiltonian Eq.~\eqref{eq: app H gamma} is non-interacting, the free energy can be readily computed: 
\begin{equation}
F(\vecg{\Delta})= -\frac{2}{\beta}\sum_\vk\ln (1+e^{-\beta E_\vk}) + \sum_\vk (\xi_\vk-E_\vk)+K_0.
\end{equation}
For convenience, it is useful to consider instead the condensation energy density, defined as the free energy density difference from the normal state, which reads
\begin{equation}\label{eq: app Fcond}
    f_{\text{cond}}(\vDelta)\equiv\frac{F(\vecg{\Delta})-F(0)}{\mathcal{V}} =k_0+k_1,
\end{equation}
where
\begin{align}
    k_0&\equiv \frac{K_0}{\mathcal{V}} = \sum_{\gamma\delta}\Delta^*_\gamma V_{\gamma\delta}^{-1} \Delta_\delta,\label{eq: app K0}\\
    k_1&=\frac{1}{\mathcal{V}}\sum_\vk \left[-\frac{2}{\beta}\ln \left(\frac{1+e^{-\beta E_\vk}}{1+e^{-\beta|\xi_\vk|}}\right)+\left(|\xi_\vk|-E_\vk  \right) \right]\label{eq: app K1}.
\end{align}
By taking the continuum limit, this expression becomes suitable for numerical computations of the free energy. Minimizing the free energy then allows us to construct the mean-field phase diagram shown in Fig.~\ref{fig: phase diagram}(a).

 The Landau free energy is derived by expanding the mean-field free energy to fourth order in $\Delta_\alpha$. For the term $k_1$, we obtain
\begin{align}
k_1&=\frac{1}{\mathcal{V}}\sum_\vk \left[-\frac{|\Delta_\vk|^2}{2\xi_\vk}\tanh\left(\frac{\beta \xi_\vk}{2} \right) +\frac{-\beta\xi_\vk+\sinh(\beta\xi_\vk)}{8\xi_\vk^3(1+\cosh(\beta\xi_\vk))}|\Delta_\vk|^4+\mathcal{O}(|\Delta_\vk|)^6\right]\nonumber\\
&\simeq -N(0)\underbrace{\int_{-\Lambda}^\Lambda \d \xi_\vk \frac{1}{2\xi_\vk}\tanh\left(\frac{\beta \xi_\vk}{2} \right)}_{\to\ln \left(\frac{2e^\gamma}{\pi}\frac{\Lambda}{k_BT} \right)}\int \frac{\d\Omega_\vk}{4\pi}|\Delta_\vk|^2+ N(0)\underbrace{\int_{-\Lambda}^\Lambda \d\xi_\vk  \frac{-\beta\xi_\vk+\sinh(\beta\xi_\vk)}{8\xi_\vk^3(1+\cosh(\beta\xi_\vk))}}_{\to \frac{7\zeta(3)}{(4\pi k_BT)^2}}\int \frac{\d\Omega_\vk}{4\pi}|\Delta_\vk|^4\nonumber\\
&=-\sum_{\alpha} N(0)\ln \left(\frac{2e^\gamma}{\pi}\frac{\Lambda}{k_BT} \right)|\Delta_\alpha|^2+\sum_{\alpha\beta\gamma\delta}\frac{7\zeta(3)}{16 \pi^2}\frac{N(0)}{(k_BT)^2}\int \frac{\d\Omega_\vk}{4\pi}\phi_\alpha^*(\vk)\phi_\beta^*(\vk)\phi_\gamma(\vk)\phi_\delta(\vk)\Delta_\alpha^*\Delta_\beta^*\Delta_\gamma\Delta_\delta.
\end{align}
Combining with Eqs.~\eqref{eq: app Fcond} and \eqref{eq: app K0}, we obtain the free energy in Eq.~\eqref{eq: Landau free energy} of the main text.

\subsection{II. Time-dependent Ginzburg-Landau theory}
In this supplemental material, we provide a detailed derivation of the TDGL equation [Eq.~\eqref{eq: TDGL} of the main text] for the model defined in Eq.~\eqref{eq: Ham}.

\subsubsection{A. Nonequilibrium Green's functions}

The dynamics of our system is governed by the mean-field Hamiltonian
\begin{equation}
    \hat{H}_{\text{MF}}=\hat{H}_0 +\hat{H}_{\Delta}(t),
\end{equation}
where $\hat{H}_0 = \sum_{\vk,\tau \tau^\prime} \hat{\Psi}^\dagger_{\vk\tau} h_{\vk,\tau\tau^\prime} \hat{\Psi}_{\vk\tau^\prime}$ is the kinetic term and $\hat{H}_\Delta(t) = \sum_{\vk,\tau \tau^\prime} \hat{\Psi}^\dagger_{\vk\tau} W_{\vk,\tau\tau^\prime} (t)\hat{\Psi}_{\vk\tau^\prime}$ is the pairing Hamiltonian. Here $\hat{\Psi}_\vk=(\hat{c}_{\vk\up},\hat{c}^\dagger_{-\vk\down})^T$ is the Nambu spinor and $\tau=p,h$ the particle-hole index. The matrix kernels are
\begin{align}
    h_\vk&=\begin{bmatrix}
        \xi_\vk&0\\
        0&-\xi_\vk
    \end{bmatrix},& W_\vk(t)&=\begin{bmatrix}
        0&-\Delta_\vk(t)\\
        -\Delta_\vk^*(t)&0
    \end{bmatrix},
\end{align}
where the gap function is given by
\begin{equation}\label{eq: gap pair amplitude}
    \Delta_\vk(t)=\frac{1}{\mathcal{V}}\sum_{\vkp}V_{\vk\vkp}(t) \expect{\hat{c}_{-\vk\down}(t)\hat{c}_{\vk\up}(t)}.
\end{equation}
As the gap function is small near the phase transition, we treat it as a perturbation. We suppose that in the distant past, the interaction potential is zero and is switched on adiabatically (but not switched off). This amounts to introducing a factor $e^{\delta t}$, with positive and infinitesimal $\delta$, to $V_{\vk\vkp}(t)$ and $\Delta(t)$. This factor is a mathematical trick to ensure convergence and will be omitted unless explicitly required.

\begin{figure}[t]
\includegraphics[width=0.8\columnwidth]{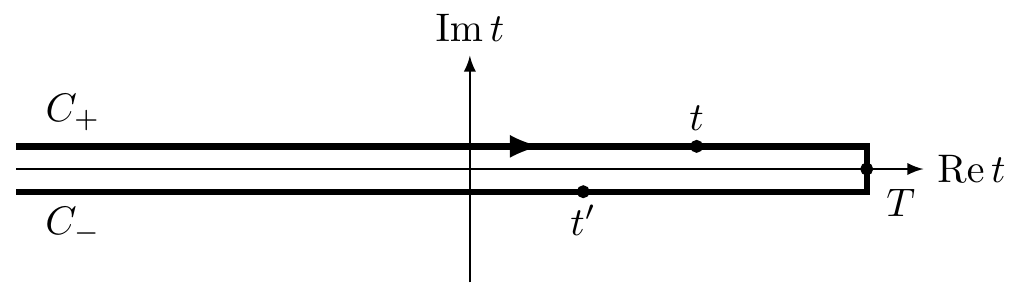}
\caption{\label{fig: Keldysh}The Keldysh contour $C$ consists of two branches that run along the real axis: the forward branch $C_+$ and the backward branch $C_-$. The points $t$ and $t^\prime$ represent field insertions along the contour. The time $T>t,t^\prime$ is when the contour changes direction and is taken to be in the limit $T\to \infty$.}
\end{figure}

Because the system is not in equilibrium, we employ contour-ordered Green's functions to facilitate the perturbative expansion. Suppose in the distant past, when the gap function is zero, the system is thermal with density matrix $\hat{\rho}_0=e^{-\beta\hat{H}_0}/Z_0$, where $Z_0=\Tr e^{-\beta\hat{H}_0}$ is the free partition function. The contour-ordered Nambu-Gorkov Green's function is defined as
\begin{align}
    \mathcal{G}_{\vk,\tau\tau^\prime}(t,t^\prime)&\equiv -i\expect{T_C \hat{\Psi}_{\vk\tau}(t)\hat{\Psi}^\dagger_{\vk\tau^\prime}(t^\prime)}\\
    &
    =-i\begin{bmatrix}
         \expect{T_C \hat{c}_{\vk\up}(t)\hat{c}^\dagger_{\vk\up}(t^\prime)}& \expect{T_C \hat{c}_{\vk\up}(t)\hat{c}_{-\vk\down}(t^\prime)}\\
         \expect{T_C \hat{c}^\dagger_{-\vk\down}(t)\hat{c}^\dagger_{\vk\up}(t^\prime)}&  \expect{T_C \hat{c}_{-\vk\down}^\dagger(t)\hat{c}_{-\vk\down}(t^\prime)}
    \end{bmatrix}
    \equiv \begin{bmatrix}
        G_{\vk}(t,t^\prime)&F_\vk(t,t^\prime)\\
        \bar{F}_{\vk}(t,t^\prime)&\bar{G}_{\vk}(t,t^\prime)
    \end{bmatrix}.
\end{align}
The times $t,t^\prime$ are defined on the Keldysh contour $C$, shown in Fig.~\ref{fig: Keldysh}, which consists of a forward branch $C_+$ and a backwards branch $C_-$. The contour-ordering operator, $T_C$, orders times later in the contour on the left. Here the thermal expectations are taken with respect to the thermal state in the distant past, $\hat{\rho}_0$. The components of the Nambu-Gorkov Green's function are not independent but satisfy $F^*_\vk(t,t^\prime)= -\bar{F}_\vk(t^\prime,t)$ and $G_\vk^*(t,t^\prime)=-\bar{G}_{-\vk}(t^\prime,t)$.

The contour-ordered Green's function is related to some more familiar real time Green's functions. Specifically,
\begin{equation}
    \mathcal{G}_{\vk,\tau\tau^\prime} (t,t^\prime) = \begin{cases}
        \mathcal{G}^T_{\vk,\tau\tau^\prime}(t,t^\prime)\equiv -i\expect{T_t \hat{\Psi}_{\vk\tau}(t)\hat{\Psi}^\dagger_{\vk\tau^\prime}(t^\prime)}&t,t^\prime\in C_+,\\
    \mathcal{G}^<_{\vk,\tau\tau^\prime}(t,t^\prime)\equiv i\expect{\hat{\Psi}^\dagger_{\vk\tau^\prime}(t^\prime)\hat{\Psi}_{\vk\tau}(t)}&t\in C_+, t^\prime\in C_-,\\
        \mathcal{G}^>_{\vk,\tau\tau^\prime}(t,t^\prime)\equiv -i\expect{\hat{\Psi}_{\vk\tau}(t)\hat{\Psi}^\dagger_{\vk\tau^\prime}(t^\prime)}&t\in C_-,t^\prime \in C_+,\\
          \mathcal{G}^{\tilde{T}}_{\vk,\tau\tau^\prime}(t,t^\prime)\equiv -i\expect{\tilde{T}_t \hat{\Psi}_{\vk\tau}(t)\hat{\Psi}^\dagger_{\vk\tau^\prime}(t^\prime)}&t,t^\prime\in C_-.
    \end{cases}
\end{equation}
Here $\mathcal{G}^T_\vk(t,t^\prime)$, and $\mathcal{G}^{\tilde{T}}_\vk(t,t^\prime)$ are the time-ordered and anti-time-ordered Green's functions, respectively, where $T_t$ ($\tilde{T}_t$) is the time-ordering (anti-time-ordering) operator, and
$\mathcal{G}^>_\vk(t,t^\prime)$ and $\mathcal{G}^<_\vk(t,t^\prime)$ are the greater and lesser Green's functions. For convenience, it is useful to introduce the retarded and advanced Green's functions, defined as
\begin{align}
    \mathcal{G}^R_\vk(t,t^\prime)&= \theta(t-t^\prime)\left[\mathcal{G}^>_\vk(t,t^\prime)-\mathcal{G}^<_\vk(t,t^\prime)\right],\\
    \mathcal{G}^A_\vk(t,t^\prime)&= -\theta(t^\prime-t)\left[\mathcal{G}^>_\vk(t,t^\prime)-\mathcal{G}^<_\vk(t,t^\prime)\right].
\end{align}
It can be readily verified that the Green's functions satisfy
\begin{align}
\mathcal{G}^R&=\mathcal{G}^T-\mathcal{G}^<=\mathcal{G}^>-\mathcal{G}^{\tilde{T}},\label{eq: GR relations}\\
\mathcal{G}^A&=\mathcal{G}^T-\mathcal{G}^>=\mathcal{G}^<-\mathcal{G}^{\tilde{T}}\label{eq: GA relations}.
\end{align}
The pair amplitude can be expressed in terms of the greater and lesser Green's functions: $\expect{\hat{c}_{-\vk\down}(t) \hat{c}_{\vk\up}(t)}=-iF^>_\vk(t,t)=-iF^<_\vk(t,t)$. Hence, Eq.~\eqref{eq: gap pair amplitude} can be written, in the continuum limit, as
\begin{align}\label{eq: gap equation F}
    \Delta_\vk(t)&=-i\int \frac{\d\Omega_\vk}{4\pi}N(0)V_{\vk\vkp}\int_{-\Lambda}^\Lambda \d\xi_\vkp\left[\frac{F^{>}_\vkp(t,t)+ F^{<}_\vkp(t,t)}{2} \right].
\end{align}
Thus, determining the dynamics of the gap function amounts to computing the greater and lesser anomalous Green's functions. 

The \emph{raison d'\^etre} for introducing contour-ordered Green's functions is to provide a systematic framework for constructing the perturbative series in the absence of adiabaticity. In the interaction picture, the Green's function admits the following form, akin to the Gell-Mann and Low theorem in equilibrium many-body physics:
\begin{equation}
\mathcal{G}_{\vk,\tau\tau^\prime}(t,t^\prime)= -i\expect{T_C \hat{U}_C \hat{\Psi}_{\vk\tau}(t)\hat{\Psi}^\dagger_{\vk\tau^\prime}(t^\prime)},
\end{equation}
where $\hat{U}_C=\expb{-\frac{i}{\hbar}\int_C \d t_1 \hat{H}_\Delta(t_1)}$
is the time-evolution operator along the entire Keldysh contour. In this expression, and from here onwards, all operators are in the interaction picture. Expanding the time-evolution operator and applying Wick's theorem, we obtain Dyson's equation
 \begin{align}
    \mathcal{G}_\vk(t,t^\prime) = \mathcal{G}^{(0)}_\vk(t,t^\prime)+ \frac{1}{\hbar}\int_C \d t_1  \mathcal{G}^{(0)}_\vk(t,t_1)W_{\vk}(t_1) \mathcal{G}_\vk(t_1,t^\prime),
 \end{align}
where $\mathcal{G}^{(0)}_\vk(t,t^\prime)$ is the free Green's function. By iterating Dyson's equation, we can obtain the correction to the Green's function at any order. For example, the $n$th order correction is
\begin{equation}\label{eq: n'th order correction}
    \mathcal{G}^{(n)}_\vk (t,t^\prime) = \frac{1}{\hbar^n}\int_C \d t_1 \d t_2\dots \d t_n \mathcal{G}^{(0)}_{\vk}(t,t_1)W_{\vk}(t_1)\mathcal{G}^{(0)}_{\vk}(t_1,t_2)W_\vk(t_2) \dots \mathcal{G}^{(0)}_{\vk}(t_{n-1},t_{n})W_\vk(t_n)\mathcal{G}^{(0)}_\vk(t_{n},t^\prime).
\end{equation}

To proceed, we need an explicit form for the free Green's function. As the Hamiltonian is non-interacting, the Green's function can be computed easily and is diagonal in Nambu space:
\begin{align}\label{eq: G^{(0)} diagonal}
    \mathcal{G}^{(0)}_\vk(t,t^\prime)= \begin{bmatrix}
        G_{p,\vk}(t,t^\prime)&\\
        &G_{h,\vk}(t,t^\prime)
    \end{bmatrix}.
\end{align}
For our purposes, we only require the retarded and advanced Green's functions,
\begin{align}
    G^R_{p,\vk}(t,t^\prime)&=-i\theta(t-t^\prime)e^{-i\xi_\vk t/\hbar},& G^R_{h,\vk}(t,t^\prime)&=-i\theta(t-t^\prime)e^{i\xi_\vk t/\hbar},\\
    G^A_{p,\vk}(t,t^\prime)&=i\theta(t^\prime-t) e^{-i\xi_\vk t/\hbar},& G^A_{h,\vk}(t,t^\prime)&=i\theta(t^\prime-t) e^{i\xi_\vk t/\hbar}.
\end{align}
These are the usual real-time equilibrium Green's functions associated with the Hamiltonian $h_\vk$. In the unperturbed case, the Hamiltonian is invariant under translations in time, so it is convenient to switch to the Fourier representation. Defining the Fourier transform $\mathcal{G}^{(0)}_\vk(\omega)= \int_{-\infty}^\infty \d t e^{i\omega(t-t^\prime)}\mathcal{G}_\vk^{(0)}(t-t^\prime)$, we obtain
\begin{align}
    G^R_{p,\vk}(\omega)&=\frac{\hbar}{\hbar\omega-\xi_\vk+i\epsilon},& G^R_{h,\vk}(\omega)&=\frac{\hbar}{\hbar\omega+\xi_\vk+i\epsilon},\\
    G^A_{p,\vk}(\omega)&=\frac{\hbar}{\hbar\omega-\xi_\vk-i\epsilon},& G^A_{h,\vk}(\omega)&=\frac{\hbar}{\hbar\omega+\xi_\vk-i\epsilon},
\end{align}
where $\epsilon>0$ is a positive infinitesimal. Using Eqs.~\eqref{eq: n'th order correction} and \eqref{eq: G^{(0)} diagonal}, we can deduce that to third order, the anomalous Green's function is,
\begin{equation}
    F_\vk(t,t^\prime)\simeq F^{(1)}_\vk(t,t^\prime)+ F^{(3)}_{\vk}(t,t^\prime),
\end{equation}
where
\begin{align}
    F^{(1)}_{\vk}(t,t^\prime)&= -\frac{1}{\hbar}\int_C \d t_1 G_{p,\vk}(t,t_1)\Delta_\vk(t_1)G_{h,\vk}(t_1,t^\prime),\\
    F^{(3)}_\vk(t,t^\prime)&=-\frac{1}{\hbar^3}\int_C \d t_1\d t_2 \d t_3 G_{p,\vk}(t,t_1)\Delta_\vk(t_1)G_{h,\vk}(t_1,t_2)\Delta^*_\vk(t_2)G_{p,\vk}(t_2,t_3)\Delta_\vk(t_3)G_{h,\vk}(t_3,t^\prime).
\end{align}
In the subsequent sections we will evaluate these two contributions.

\subsubsection{B. Linear term}

For the linear term, we recast the contour integral as
\begin{align}
    -\hbar F^{(1)<}_\vk(t,t^\prime)&=\int_{-\infty}^\infty \d t_1 G^T_{p,\vk}(t,t_1)\Delta_\vk(t_1)G^<_{h,\vk}(t_1,t^\prime)+\int_{\infty}^{-\infty}\d t_1 G^<_{p,\vk}(t,t_1)\Delta_\vk(t_1)G^{\tilde{T}}_{h,\vk}(t_1,t^\prime)\nonumber\\
    &=\int_{-\infty}^\infty \d t_1 \left[ G^T_{p,\vk}(t,t_1)\Delta_\vk(t_1)G^<_{h,\vk}(t_1,t^\prime)-G^<_{p,\vk}(t,t_1)\Delta_\vk(t_1)G^{\tilde{T}}_{h,\vk}(t_1,t^\prime)\right]\nonumber\\
    &=\int_{-\infty}^\infty \d t_1 \left[ G^R_{p,\vk}(t,t_1)\Delta_\vk(t_1)G^<_{h,\vk}(t_1,t^\prime)+G^<_{p,\vk}(t,t_1)\Delta_\vk(t_1)G^{A}_{h,\vk}(t_1,t^\prime)\right]\label{eq: general Langreth}.
\end{align}
In the last step, we have used Eqs.~\eqref{eq: GR relations} and \eqref{eq: GA relations}. An analogous statement can be made when $t\in C_-$ and $t^\prime\in C_+$, in which case one simply replaces the lesser functions with greater functions. This result is a slight generalization of the familiar Langreth rule~\cite{HaugJauho}
\begin{equation}
    \int_C \d t_1 A(t,t_1)B(t_1,t^\prime)=\int_{-\infty}^\infty \d t_1 \left[ A^R(t,t_1)B^<(t_1,t^\prime)+A^<(t,t_1)B^A(t_1,t^\prime)\right],
\end{equation}
when $t\in C_+$ and $t^\prime\in C_-$. This rule is more commonly and compactly written as
\begin{equation}\label{eq: Langreth 1}
    (AB)^{>(<)}= A^{R}B^{>(<)}+A^{>(<)}B^A.
\end{equation}
The left side is a contour integral and the right is one over the real axis (along with summation over any internal indices). Eq.~\eqref{eq: general Langreth} reduces to this Langreth rule when $\Delta_\vk(t_1)$ is constant in time.

If the gap function does not vary rapidly in time, we only need to account for the low-frequency modes. Taylor expanding $\Delta_\vk(t_1)$ about $t$ to first order in time,
\begin{equation}
    \Delta_\vk(t_1)\simeq \Delta_\vk(t)+\totald{\Delta_\vk}{t_1}\bigg|_t (t_1-t) e^{\delta t_1}.
\end{equation}
We have reinserted the adiabatic switch-on factor $e^{\delta t}$ in the time derivative term to ensure convergence when evaluating integrals later. Expanding the Green's functions in Fourier modes, we obtain
\begin{align}
    -\hbar F^{(1)<}_\vk(t,t^\prime) &\simeq \Delta_\vk(t)\int \frac{\d\omega}{2\pi}e^{-i\omega(t-t^\prime)} \left(G^R_{p,\vk}(\omega)G^<_{h,\vk}(\omega)+G^<_{p,\vk}(\omega)G^A_{h,\vk}(\omega) \right)\nonumber\\
    &+i\totald{\Delta_\vk}{t}(t) \int \frac{\d\omega}{2\pi}e^{-i\omega(t-t^\prime)}\left[\left(\partial_\omega G^R_{p,\vk}(\omega)\right)G^<_{h,\vk}(\omega-i\delta)+\left(\partial_\omega G^<_{p,\vk}(\omega)\right)G^A_{h,\vk}(\omega-i\delta) \right].
\end{align}
A similar expression can be obtained for the greater function by replacing all the lesser functions with greater ones. Using the identity
\begin{equation}\label{eq: fluctuation dissipation}
    \mathcal{G}^{(0)>}_\vk(\omega)+\mathcal{G}^{(0)<}_\vk(\omega)=\tanh\left( \frac{\beta \hbar\omega}{2}\right)\left(\mathcal{G}^{(0)R}_\vk(\omega)-\mathcal{G}^{(0)A}_\vk(\omega) \right), 
\end{equation}
which is valid for any equilibrium Green's function, we obtain
\begin{equation}
    \frac{F^{(1)>}_\vk(t,t^\prime)+F^{(1)<}_\vk(t,t^\prime)}{2} \simeq -\frac{\Delta_\vk(t)}{\hbar} K^{(0)}_\vk(t,t^\prime)-\frac{i}{\hbar}\totald{\Delta_\vk}{t}(t)K_\vk^{(1)}(t,t^\prime),
\end{equation}
where
\begin{align}
    K^{(0)}_\vk(t,t^\prime)&=\frac{1}{2}\int \frac{\d\omega}{2\pi}e^{-i\omega(t-t^\prime)}\tanh\left(\frac{\beta\hbar\omega}{2} \right) \left[G^R_{p,\vk}(\omega)G^R_{h,\vk}(\omega)-G^A_{p,\vk}(\omega)G^A_{h,\vk}(\omega) \right],\\
    K^{(1)}_\vk(t,t^\prime)&=\frac{1}{2}\int \frac{\d\omega}{2\pi}e^{-i\omega(t-t^\prime)}\bigg\{\tanh\left(\frac{\beta\hbar\omega}{2} \right) \bigg[ \left(\partial_\omega G^R_{p,\vk}(\omega)\right)G^R_{h,\vk}(\omega-i\delta)-\left(\partial_\omega G^A_{p,\vk}(\omega)\right)G^A_{h,\vk}(\omega-i\delta) \bigg]\nonumber\\
    &\hspace{3cm}+\partial_\omega\tanh\left(\frac{\beta\hbar\omega}{2} \right)  \left[ G^R_{p,\vk}(\omega)- G^A_{p,\vk}(\omega)\right]G^A_{h,\vk}(\omega-i\delta)\bigg\}.
\end{align}

The functions $K^{(0)}_\vk(t,t^\prime)$ and $K^{(1)}_\vk(t,t^\prime)$ can be evaluated by contour integration. Without loss of generality, suppose $t^\prime = t^-$, where $t^-$ denotes infinitesimally smaller than $t$. In this case, we close the integral path with a semicircular arc of infinite radius at the bottom half of the complex plane. Applying the residue theorem, we obtain
\begin{align}
    K^{(0)}(t,t^-)&= -\frac{i\hbar}{2\xi_\vk}\tanh\left(\frac{\beta\xi_\vk}{2}\right),\\
    K^{(1)}(t,t^-)&=\frac{-i\hbar^2}{(2\xi_\vk-i\delta)^2}\tanh\left(\frac{\beta\xi_\vk}{2} \right).
\end{align}
Therefore, the first-order correction to the pair amplitude is
\begin{equation}
     \frac{F^{(1)>}_\vk(t,t^\prime)+F^{(1)<}_\vk(t,t^\prime)}{2} \simeq \frac{i\Delta_\vk(t)}{2\xi_\vk}\tanh\left(\frac{\beta\xi_\vk}{2} \right)- \totald{\Delta_\vk}{t}(t)\frac{\hbar}{(2\xi_\vk-i\delta)^2}\tanh\left(\frac{\beta\xi_\vk}{2} \right).
\end{equation}
Integrating over $\xi_\vk$,
\begin{equation}\label{eq: first order}
    \int_{-\Lambda}^\Lambda \d\xi_\vk \left[\frac{F^{(1)>}_\vk(t,t^\prime)+F^{(1)<}_\vk(t,t^\prime)}{2} \right]=i\ln \left(\frac{2e^\gamma}{\pi}\frac{\Lambda}{k_BT} \right)\Delta_\vk(t) -i\frac{\pi\hbar}{8\pi k_BT}\totald{\Delta_\vk}{t}(t).
\end{equation}

\subsubsection{C. Cubic term}
We now turn to to the cubic term. As near the phase transition, the gap is already small, we can assume $\Delta_\vk(t_1)\simeq\Delta_\vk(t)$. Hence, the cubic contribution is
\begin{equation}
-\hbar^3F^{(3)}_\vk(t,t^\prime)=|\Delta_\vk(t)|^2\Delta_\vk(t)\int_C \d t_1\d t_2 \d t_3 G_{p,\vk}(t,t_1)G_{h,\vk}(t_1,t_2)G_{p,\vk}(t_2,t_3)G_{h,\vk}(t_3,t^\prime).
\end{equation}
As this is a straightforward convolution of four Green's functions, we can directly apply the usual Langreth rules. Combining the rules~\eqref{eq: Langreth 1} and
\begin{equation}
    (AB)^{R(A)}= A^{R(A)}B^{R(A)},
\end{equation}
we can iterate to obtain the contour convolution of four functions:
\begin{align}
	(ABCD)^>&=A^R(BCD)^>+A^>(BCD)^A\nonumber\\
	&=A^RB^R(CD)^>+A^RB^>(CD)^A+A^>B^AC^AD^A\nonumber\\
	&=A^RB^RC^RD^>+A^RB^RC^>D^A+A^RB^>C^AD^A+A^>B^AC^AD^A.
\end{align}
Switching to the Fourier representation and using the identity Eq.~\eqref{eq: fluctuation dissipation}, we obtain
\begin{align}
    &\frac{F_\vk^{(3)>}(t,t^\prime)+F_\vk^{(3)<}(t,t^\prime)}{2}\nonumber
    \\
    &= -\frac{|\Delta_\vk(t)|^2 \Delta_\vk(t)}{\hbar^3}\int \frac{\d\omega}{2\pi}e^{-i\omega(t-t^\prime)}\left[ 	G_{p,\vk}^R(\omega)G_{h,\vk}^R(\omega)G_{p,\vk}^R(\omega)G_{h,\vk}^R(\omega)-(R\leftrightarrow A)\right]\tanh\left(\frac{\beta\hbar\omega}{2} \right).
\end{align}
Evaluating the frequency integral with the same prescription as before, we obtain
\begin{align}
    \frac{F_\vk^{(3)>}(t,t)+F_\vk^{(3)<}(t,t)}{2}=i|\Delta_\vk(t)|^2|\Delta_\vk(t)\left[-\frac{1}{4\xi_\vk^3}\tanh\left(\frac{\beta\xi_\vk}{2} \right)+\frac{\beta}{8\xi_\vk^2}\sech^2\left(\frac{\beta\xi_\vk}{2} \right)\right].
\end{align}
Finally, performing the $\xi_\vk$ integral,
\begin{equation}\label{eq: third order}
    \int_{-\Lambda}^\Lambda \d\xi_\vk \left[\frac{F_\vk^{(3)>}(t,t)+F_\vk^{(3)<}(t,t)}{2}\right]=-i\frac{7\zeta(3)}{8\pi^2}\frac{1}{(k_BT)^2}|\Delta_\vk(t)|^2\Delta_\vk(t).
\end{equation}

\subsubsection{D. Gap equation}
We can now combine all our results to obtain the TDGL equation. Substituting Eqs.~\eqref{eq: first order} and \eqref{eq: third order} into Eq.~\eqref{eq: gap equation F}, we obtain
\begin{align}
    \Delta_\vk(t)= \int \frac{\d\Omega_\vk}{4\pi}N(0)V_{\vk\vkp}\left[\ln \left(\frac{2e^\gamma}{\pi}\frac{\Lambda}{k_BT} \right)\Delta_\vkp(t)-\frac{\pi\hbar}{8k_BT} \totald{\Delta_\vkp}{t}(t)-\frac{7\zeta(3)}{8\pi^2}\frac{1}{(k_BT)^2}|\Delta_\vkp(t)|^2\Delta_\vkp(t) \right].
\end{align}
Using Eqs.~\eqref{eq: partial wave decomposition} and \eqref{eq: Delta partial wave decomposition}, the order parameter coefficients are given by
\begin{align}
    \Delta_\alpha = \sum_\beta N(0)V_{\alpha\beta}\left[\ln \left(\frac{2e^\gamma}{\pi}\frac{\Lambda}{k_BT} \right)\Delta_\beta -  \frac{\pi\hbar}{8k_BT}\totald{\Delta_\beta}{t}-\frac{7\zeta(3)}{8\pi^2}\frac{1}{(k_BT)^2}\int \frac{\d\Omega_{\vkp}}{4\pi}\phi_\beta^*(\vkp)|\Delta_{\vkp}|^2\Delta_{\vkp}\right].
\end{align}
Rearranging, we obtain the TDGL equation~\eqref{eq: TDGL} with $\eta=\pi N(0)/(8 k_BT)$.

\section{III. Relaxation of the superconducting gap function}
In this supplemental material, we examine the relaxation of the superconducting gap function when slightly perturbed from the thermodynamic ground state. In particular, we show that the phase $\varphi$ (which we denote as $\varphi_-$ below) is unchanged under the relaxation process.

Let us first obtain the equations of motion for the fluctuations in the order parameter. Consider a small deviation from the equilibrium configuration,
\begin{equation}\label{eq: deviations}
    \vDelta(\vecg{\lambda})= (\Delta_0+\delta\Delta_-)e^{i\varphi_-}\hat{\vecg{\Delta}}_-(\vlambda)+\delta\Delta_+e^{i\varphi_+}\hat{\vecg{\Delta}}_+(\vlambda),
\end{equation}
where $\Delta_0 = \sqrt{-a_-/b_-}$. For times $t\sim \tau_-\ll \tau$, the parameters $\vlambda$ can be treated as constant. Substituting Eq.~\eqref{eq: deviations} into the TDGL equation~\eqref{eq: TDGL}, the time derivative term becomes
\begin{align}
    \partial_t \Delta_\alpha\simeq(\partial_t \delta \Delta_-) e^{i\varphi_-}\hat{\Delta}_{-,\alpha}+i\Delta_0 (\partial_t\varphi_-) e^{i\varphi_-}\hat{\Delta}_{-,\alpha}  + (\partial_t \delta \Delta_+) e^{i\varphi_+}\hat{\Delta}_{+,\alpha} +i\delta\Delta_+ (\partial_t \varphi_+) e^{i\varphi_+}\hat{\Delta}_{+,\alpha}.
\end{align}
Similarly, expanding the linear term,
\begin{align}
    \sum_\beta A_{\alpha\beta}\Delta_\beta= a_- (\Delta_0+\delta\Delta_-)e^{i\varphi_-}\hat{\Delta}_{-,\alpha}+a_+ \delta\Delta_+ e^{i\varphi_+}\hat{\Delta}_{+,\alpha}.
\end{align}
Finally, the cubic term,
\begin{align}
\sum_{\beta\gamma\delta}B_{\alpha\beta\gamma\delta}\Delta^*_\beta\Delta_\gamma\Delta_\delta &\simeq \hat{\Delta}_{-,\alpha}\Delta_0^2\left[b_-(\Delta_0+3\delta\Delta_-) e^{i\varphi_-}+2\tilde{B}_{---+}\delta \Delta_+ e^{i\varphi_+}+\tilde{B}_{-+--}\delta\Delta_+ e^{i(2\varphi_--\varphi_+)} \right]\nonumber\\
&+\hat{\Delta}_{+,\alpha}\Delta_0^2\left[\tilde{B}_{+---}(\Delta_0+3\delta\Delta_-)e^{i\varphi_-}+2\tilde{B}_{+--+}\delta\Delta_+ e^{i\varphi_+}+\tilde{B}_{++--}\delta\Delta_+ e^{i(2\varphi_--\varphi_+)} \right].
\end{align}
For convenience, we denote $\tilde{B}_{abcd}=\sum_{\alpha\beta\gamma\delta}B_{\alpha\beta\gamma\delta}\hat{\Delta}_{a,\alpha}^*\hat{\Delta}_{b,\beta}^*\hat{\Delta}_{c,\gamma}\hat{\Delta}_{\delta}$, with $a,b,c,d=\pm$, as the quartic coefficients in the $\hat{\vecg{\Delta}}_{\pm}$ basis and $b_-\equiv \tilde{B}_{----}$. Substituting in these expansions, the TDGL equation in the regime $a_+\gg |B_{\alpha\beta\gamma\delta}||\Delta_0|^2$ becomes
\begin{align}
    -\hbar\eta \left(\partial_t \delta\Delta_-+i\Delta_0\partial_t\varphi_-\right)&\simeq -3a_-\delta\Delta_-+\Delta_0^2\left[2b_{+-}\delta\Delta_+ e^{i(-\psi+\varphi_+-\varphi_-)}+b_{+-}\delta\Delta_+ e^{i(\psi+\varphi_--\varphi_+)} \right]\label{eq: Delta_-},\\
    -\hbar\eta \left(\partial_t \delta\Delta_++i\Delta_0\partial_t\varphi_+\right)&\simeq a_+\delta\Delta_++ 3\Delta_0^2 b_{+-}\delta\Delta_- e^{i(\psi+\varphi_--\varphi_+)},\label{eq: Delta_+}
\end{align}
where we parameterized $\tilde{B}_{+---}=b_{+-}e^{i\psi}$ and used $\tilde{B}_{+---}=\tilde{B}_{----+}^*=\tilde{B}_{-+--}$.

The equations \eqref{eq: Delta_-} and \eqref{eq: Delta_+} can be solved by adiabatic elimination. Because $\tau_+\ll \tau_-$, the variables $\delta \Delta_+$ and $\varphi_+$ quickly relax to quasisteady state. For times $t\sim \tau_-$,
\begin{align}
    \delta \Delta_+ &\simeq -\frac{3\Delta_0^2b_{+-}}{a_+}\delta \Delta_- \cos(\psi+\varphi_--\varphi_+),\\
    0&\simeq \sin(\psi+\varphi_--\varphi_+).
\end{align}
The only stable solution to the latter equation is $\varphi_+ = \psi+\varphi_-$. Hence,
\begin{align}
    \delta \Delta_+ e^{i \varphi_+}\simeq -\frac{3\Delta_0^2b_{+-}}{a_+}\delta \Delta_- e^{i(\psi+\varphi_-)}.
\end{align}
Substituting into Eq.~\eqref{eq: Delta_-}, we obtain
\begin{align}
    -\hbar\eta\left(\partial_t \delta\Delta_- + i\Delta_0 \partial_t \varphi_- \right) 
    &\simeq -3a_-^\prime \delta\Delta_-,
\end{align}
where $a_-^\prime = a_- + 3\Delta_0^2b_{+-}^2/a_+$. This implies $\delta \Delta_-$ decays exponentially with characteristic timescale $\hbar \eta/(3 |a_-^\prime|)\sim \tau_-$ and the phase remains constant, as asserted.

\end{document}